%% file: spec.tex
\documentclass[twoside]{sfm}

\input{sf.def}

\begin{document}

\input{kochukhov/kochukhov.tex}

\end{document}

%% file: kochukhov/kochukhov.tex
\pagebreak

\thispagestyle{titlehead}

\setcounter{section}{0}
\setcounter{figure}{0}
\setcounter{table}{0}

\markboth{Kochukhov O.}{Magnetic fields in A stars besides Ap stars}

\titl{Magnetic fields in A stars besides Ap stars}{Kochukhov O.}
{Department of Physics and Astronomy, Uppsala University, Sweden \\ email: {\tt oleg.kochukhov@physics.uu.se}}

\abstre{
I review ongoing efforts to understand the incidence of magnetism in intermediate-mass stars that are different from the magnetic Ap stars. This includes the search for magnetic fields in chemically peculiar stars of the Am and HgMn types as well as in normal A and late-B stars. I discuss different techniques for detection of weak stellar magnetic fields and present a critical evaluation of the recent magnetic detections in non-Ap stars. Special attention is given to the magnetic status of HgMn stars and to the discovery of weak polarization signatures in Sirius and Vega.
}

\baselineskip 12pt

\section{Introduction}

Stellar magnetism studies carried out during six decades since the discovery \cite{babcock47} of global magnetic fields in peculiar A stars have firmly established the bimodal character of the incidence of magnetic fields among intermediate-mass main sequence stars. On the one hand, the majority of these stars lack fields exceeding several hundred G, and are rapid rotators with approximately solar chemical composition. On the other hand, certain sub-groups of chemically peculiar (CP) stars possess globally organized magnetic fields with strength of up to $\sim$\,30~kG \cite{donati09}. These magnetic CP (or Ap/Bp) stars are characterized by slow rotation and exhibit conspicuously non-solar surface chemical abundance patterns. Among cooler CP stars magnetic fields are invariably found in the objects with SrCrEu spectral peculiarity. As the temperature increases, similar magnetic properties are observed for Si-rich and He-abnormal B-type stars. Simultaneously, there exists another group of CP stars (Am on the cooler side and HgMn/PGa on the hotter side) for which no credible evidence of  magnetic fields has ever been presented. 

Thus, there exists a magnetic dichotomy among intermediate-mass A and B stars: strongly magnetic objects share the H-R diagram with the stars deemed to be completely void of surface magnetic fields. Recent improvements in the observational techniques of stellar magnetometry supported this magnetic dichotomy paradigm, demonstrating that every SrCrEu and Si-rich Ap/Bp star has the field of at least 300~G \cite{auriere07}. This limit is physically significant because for brighter stars it significantly exceeds the sensitivity of modern spectropolarimetric surveys. Furthermore, theoretical simulations \cite{braithwaite06} suggested a convincing framework for understanding the interior structure and stability of the global magnetic fields in Ap stars, although the questions of how A and B stars acquire their fields in the first place and why only 10\% of stars do so remain unanswered. 

At the same time, a number of recent studies claimed discoveries of magnetic fields in A and B stars other than Ap stars, challenging the classical division of intermediate-mass stars into ``magnetic'' and ``non-magnetic'' groups. Many of these claims were subsequently refuted, but a few were supported by independent analyses. Here I attempt to clarify the current observational picture of the incidence of magnetic fields in A and late-B stars other than magnetic Ap stars by summarizing and critically evaluating the outcomes of relevant recent studies.

\section{Methods of detection of weak stellar magnetic fields}

\subsection{Low-resolution spectropolarimetry}

The low-resolution spectropolarimetry with FORS1/2 instruments at ESO VLT \cite{bagnulo02} is one of the most common techniques used during past decade for large-scale searches of stellar magnetic fields. This method estimates the mean longitudinal magnetic field, $\langle B_{\rm z}\rangle$, by correlating the Stokes $I$ derivative with the Stokes $V$ signal in the wings of hydrogen lines or in unresolved blends of metal lines.  FORS1/2 spectropolarimetry appears to be robust when applied to strongly magnetic Ap stars \cite{bagnulo09,kochukhov06}. 
However, FORS instrument cannot be reliably used to study magnetic fields below a few hundred G. Detailed assessment of certain controversial FORS results and re-analysis of the entire FORS1 archive revealed the presence of several artifacts related to flexures in this Cassegrain-mounted instrument \cite{bagnulo12,bagnulo13}. There are also significant ambiguities in the data reduction, leading to changes in the resulting field estimates well in excess of the formal photon noise error bars, in response to small variation of the reduction parameters. To summarize results of \cite{bagnulo12,bagnulo13}, the FORS1/2 detections of weak magnetic fields are trustworthy only if $\langle B_{\rm z}\rangle$ significance of better than 5--6$\sigma$ is obtained. But any field measurements may turn out to be spurious below 100--200~G due to occasional large systematic errors. Such detections require confirmation by other instruments.

\subsection{Moment technique}

The moment technique \cite{mathys94,mathys06} was originally introduced in the context of analysis of moderate-quality circular polarization spectra of Ap stars. In this method different magnetic field moments (mean longitudinal field, quadratic field, etc.) are inferred from the moments of Stokes $I$ and $V$ profiles of individual metal lines. While successfully applied to strongly magnetic Ap stars, the moment method has not been verified against other techniques or synthetic spectrum calculations for fields weaker than $\sim$\,100~G. Some spurious field detections obtained with this method (see discussion in \cite{kochukhov13} and below) suggest that it may suffer from hitherto unrecognized biases when applied to noisy circular polarization data.

\subsection{Least-squares deconvolution}

The least-squares deconvolution (LSD \cite{donati97,kochukhov10}) relies on combining intensity and polarization profiles of a large number of metal lines into mean Stokes profiles characterized by a very high signal-to-noise ratio. The primary field detection diagnostic is the presence of a statistically significant signature in the LSD Stokes $V$ profile. The field strength can be quantified by computing $\langle B_{\rm z}\rangle$ and other field moments from the LSD profiles. Its ability to recover a high-quality mean polarization signature represents a major advantage of LSD compared to methods that only estimate the mean longitudinal field. In particular, LSD is sensitive to complex fields \cite{kochukhov13a} and to magnetic field geometries with negligible mean longitudinal field (e.g. toroidal fields or equator-on oblique dipoles).

The LSD technique has been successfully applied to strongly-magnetic Ap stars \cite{auriere07} and to a wide range of late-type stars with different activity levels \cite{petit08}, some having sub-G magnetic fields \cite{auriere09}. The performance and limitations of LSD were thoroughly explored using synthetic Stokes spectra \cite{kochukhov10}. Thus, in comparison to the low-resolution spectropolarimetry with FORS1/2 and the moment technique, LSD is much better understood and is consequently far more reliable approach to finding weak stellar magnetic fields.

\section{Recent observational results}

\subsection{Am and normal A/B stars}

Several magnetic field surveys have addressed the question of the incidence of magnetism in Am and normal A/B-type stars. High-resolution observations with MuSiCoS \cite{shorlin02} and NARVAL \cite{auriere10} spectropolarimeters probed the presence of magnetic field in about 40 normal A and Am stars. No field detections were reported, with typical $\langle B_{\rm z}\rangle$ uncertainties of 10--50~G for most targets, but down to 1--3~G for several bright narrow-line Am stars. The FORS1 cluster survey \cite{bagnulo09} encompassed over 100 relatively faint A and B stars without noticeable spectral peculiarities, finding no field above 100--200~G. The FORS investigation of a sample of RR~Lyr pulsators \cite{kolenberg09} also yielded null results at the level of $\approx$\,30~G. 

On the other hand, magnetic field with $|\langle B_{\rm z}\rangle|$ of up to 380~G was reported for A0 supergiant HD\,92207 from FORS2 observations \cite{hubrig12a}. A subsequent study \cite{bagnulo13} demonstrated that these FORS2 spectra were affected by erratic wavelength variations happening on short time scale and that the field detection in HD\,92207 is spurious. High-precision HARPSpol measurements of this star \cite{bagnulo13} have established an upper limit of only 10~G for the mean line-of-sight magnetic field component.

\subsection{$\beta$~Cep, SPB and Be stars}

Initially, observations with FORS1 \cite{hubrig09a} suggested an unusually high incidence of weak magnetic fields in spectroscopically normal pulsating $\beta$~Cep and SPB late-B stars. However, the follow-up high-resolution studies \cite{silvester09} and re-analysis of the FORS1 data \cite{bagnulo12} could confirm only a couple of these detections. Moreover, ``magnetic field models'' of six $\beta$~Cep and SPB stars published by \cite{hubrig11} were shown to be invalid for all but one star \cite{shultz12} as the predicted Stokes $V$ profiles turned out to be many times stronger than the actual upper limit of the circular polarization signals observed for these stars. There is no doubt that a few $\beta$~Cep and SPB stars possess global dipolar-like fields \cite{neiner12,silvester09}, but the fraction of magnetic stars among late-B pulsators is not anomalously high and is generally consistent with the overall $\sim$10\% incidence of magnetism for the entire group of mid- to late-B main sequence stars.

A re-assessment of the FORS1 archive \cite{bagnulo12} also did not confirm any of the field detections reported for classical Be stars \cite{hubrig06a,hubrig09}. It was concluded that magnetic fields above 100~G rarely if ever occur in these objects. Remarkably, the MiMeS high-resolution spectropolarimetric survey \cite{wade12} has failed to detect the field in any of 58 studied Be stars. Thus, it appears that the Be and surface magnetism phenomena are mutually exclusive.

\subsection{HgMn stars}

Since the first reports of the magnetic fields in Ap stars, it was recognized that the presence of the field is often accompanied by the surface inhomogeneities in chemical abundance distribution -- starspots 
-- and that magnetic and line strengths variations occur with the same period. These observations inspired the oblique rotator model \cite{stibbs50}, according to which both the field geometry and the spot topologies are stable and the prominent periodic spectrophotometric and magnetic variability of Ap stars is attributed entirely to the changing aspect angle due to stellar rotation. The stability of the surface inhomogeneities in Ap stars is confirmed by repeatability of their photometric light curves. Apart from occasional slow down and precession, the pattern of photometric variability in these stars does not change on the time scales of at least several decades \cite{adelman01,mikulasek08}. 

The straightforward and conceptually attractive picture of the one to one correlation between starspots and magnetic fields had to be revised with the discovery of chemical inhomogeneities in HgMn stars \cite{adelman02,hubrig06a,kochukhov05,kochukhov11}. Contrary to the wide-spread belief that non-magnetic stars should have homogeneous atmospheres, it was ascertained that some HgMn stars exhibit a low-level spectrum variability, typically in the lines of strongly overabundant elements (Hg, Pt, Sr, Y). Moreover, temporal behavior of these chemical spots turned out to be noticeably different from those in magnetic Ap stars: several studies demonstrated that spots in HgMn stars change their configuration on the time scale of one year or less \cite{kochukhov07,korhonen13}.

However, all attempts to find magnetic fields that might be associated with these chemical inhomogeneities have failed to yield a single undisputed magnetic field detection. For example, a comprehensive HARPSpol survey of nearly 50 HgMn stars \cite{maka11a} and previous high-resolution spectropolarimetric studies \cite{auriere10,shorlin02} inferred upper limits of 1--10~G for $\langle B_{\rm z}\rangle$ using LSD analysis. The best precision was obtained for HgMn stars with the sharpest spectral lines, which show no detectable spectral variability. But even intense dedicated observations targeting individual HgMn stars with clear spot signatures revealed no magnetic fields \cite{folsom10,kochukhov11,maka11b,maka12,wade06}, with the best precision of 2--3~G obtained for $\mu$~Lep \cite{kochukhov11}.

Despite these results, sporadic reports of a few tens to a few hundreds G magnetic field detections have appeared in the literature \cite{hubrig06a,hubrig06b,hubrig12}. These claims were based on moment analysis of the archival HARPSpol circular polarization spectra and on observations with low- (FORS1/2 at VLT) and intermediate-resolution (SOFIN at NOT) Cassegrain mounted instruments. None of these analyses presented a direct detection of the spectral line polarization signatures for HgMn stars. Instead, the existence of the field was inferred only through non-zero $\langle B_{\rm z}\rangle$ measurements. These results have not withstood an independent scrutiny \cite{bagnulo12,folsom10,kochukhov13}. In all cases more precise high-resolution spectropolarimetric observations of the same stars and re-analysis of the archival data found no evidence of the field. As mentioned above, careful examination of the publicly available FORS1/2 data revealed instrumental artifacts and uncertainties in the reduction, rendering claims of $\le$\,100--200~G field detections with this instrument questionable. Similar instabilities may plague the low-resolution mode of the SOFIN spectropolarimeter.

To summarize, currently there exists no reliable evidence for the globally organized magnetic fields in any of the \textit{bona fide} HgMn star, including objects with chemical spots. The upper limit on possible surface fields that could still remain undetected is $\sim$\,10--30~G, although 2--3 times weaker fields have been excluded for several sharp-line stars. 

The absence of circular polarization in the line profiles does not rule out a much more complex ``tangled'' magnetic fields. Although LSD analysis of the Stokes $V$ spectra can reveal fields structured on scales down to a few degrees \cite{kochukhov13a}, one can in principle envisage even more complex turbulent fields, which only contribute to the line broadening but are invisible in polarization due to a complete cancellation of opposite field polarities.  Leaving aside the question of physical origin of such hypothetical magnetic fields, several studies tried to diagnose them in HgMn stars from high-resolution intensity spectra, using relative intensification of the spectral lines with different Zeeman splitting patterns \cite{hubrig99,hubrig01} or analyzing magnetic broadening with quadratic field diagnostic method \cite{hubrig98,hubrig12,mathys95}. Somewhat surprisingly, these analyses obtained fields of the order of 2--4~kG for a number of HgMn stars. These results appear to be in a strong contradiction with numerous detailed model atmosphere and spectrum synthesis studies of the same targets, which never required such strong fields to reproduce their Stokes $I$ spectra. This discrepancy was addressed in our recent study \cite{kochukhov13} based on detailed radiative transfer modeling of HgMn-star observations at resolving power $>10^5$. It was found that turbulent fields stronger than 200--500~G are inconsistent with spectroscopic observations of slowly rotating HgMn stars and that relative intensification and quadratic field measurements are not trustworthy as the field detection techniques due to unrealistic assumptions about magnetic line formation.

\subsection{Vega and Sirius}

An application of the LSD processing to high-resolution spectropolarimetric data recorded over a wide wavelength region enables a major gain in sensitivity to weak stellar magnetic fields. Several studies have achieved a precision of better than 1~G for $\langle B_{\rm z}\rangle$, corresponding to a polarimetric sensitivity of 10$^{-5}$ and better, for bright late-type stars \cite{auriere09}. Rapid rotation and sparse metal line spectra prevent achieving this level of precision for all but the brightest A stars, such as Vega and Sirius. For these objects a sub-G field precision can be attained by co-adding spectropolarimetric observations obtained over several nights, provided an adequate spectrometer stability. This observational methodology was exploited for Vega \cite{ligni09,petit10} and Sirius \cite{petit11} using ESPaDOnS and NARVAL instruments.

Analysis of the circular polarized LSD profiles of both stars revealed signatures with an amplitude of 10$^{-5}$ of the continuum intensity and a longitudinal field below 1~G. Detection of magnetic field in Vega was accomplished using both aforementioned instruments and was further supported by the Zeeman Doppler imaging inversions \cite{petit10}. Consistently with a narrow Stokes $V$ profile, inversions showed a relatively complex surface field structure, dominated by a polar field concentration where the field reaches 3~G locally. The short-term variation of the polarized signatures corresponds well to the rotation period expected for Vega.

On the other hand, the Stokes $V$ signature reported for Sirius \cite{petit11} defies an explanation in terms of the Zeeman effect. The mean Stokes $V$ profile of this star has a strong asymmetry between the positive and negative lobes, yielding a significant zeroth-order moment. Such Stokes $V$ profiles are known for the solar active regions characterized by strong vertical magnetic and velocity gradients \cite{sd12}. It is unknown how such exotic polarization profiles can appear in the disk-integrated flux spectrum of a star with quiescent and relatively well understood atmosphere. A possibility of persistent instrument artifact cannot be neglected, but appears unlikely given that this Stokes $V$ signature was confirmed for Sirius using a spectropolarimeter with a different design (HARPSpol, Kochukhov et al. in preparation).

In any case, observations of Vega and Sirius (if a reasonable explanation of its peculiar Stokes $V$ profile could be found) point to an entirely new manifestation of magnetism among A and B stars. These fields probably exist in all intermediate-mass stars and are weaker by about two orders of magnitude with respect to the 300~G lower limit of the Ap-star magnetic field. Clearly, more observational work is required to probe the presence of such fields in other normal bright A stars and to investigate their magnetic field topologies and evolution.

\section{Conclusions}

Numerous magnetic field surveys conducted with spectropolarimeters at large and intermediate-size telescopes have significantly increased the sample of A and B stars investigated for the presence of magnetic fields. At the same time, the literature became contaminated by spurious field detections, coming primarily from unrecognized instrumental artifacts affecting Cassegrain-mounted spectropolarimeters and from unscrupulous application of the moment technique to low signal-to-noise ratio circular polarization spectra. One should be extremely careful when interpreting these results. Recent studies showed that the LSD analysis of Stokes $V$ spectra recorded with stabilized fibre-fed spectrometers yields the least number of spurious detections and is more trustworthy.

Having these cautionary notes in mind, the following major conclusions emerge from the recent magnetic field studies of peculiar and normal A/B stars:
\begin{enumerate}
\item All Ap stars are magnetic, with a minimum dipolar strength of 300~G.
\item Weak global magnetic fields below this limit and down to 10--50~G can be excluded for all Am, HgMn, and Be stars. Tangled magnetic fields stronger than 0.2--0.5~kG are also ruled out for HgMn stars.
\item A few $\beta$~Cep and SPB stars are magnetic, but the incidence of magnetism among these B-type pulsators is not abnormally high.
\item There is a ``magnetic desert'' between 300 and $\sim$\,10~G for A stars. Below this range, Vega-like fields can exist in the majority of stars.
\end{enumerate}

\bigskip
{\it Acknowledgements.} The author is a Royal Swedish Academy of Sciences Research Fellow, supported by the grants from Knut and Alice Wallenberg Foundation and Swedish Research Council.

%% file: spec.bbl
\begin{thebibliography}{99}


\bibitem{adelman01}
{Adelman S.J., Malanushenko V., Ryabchikova T., et al.} 2001, A\&A, 375, 982

\bibitem{adelman02}
{Adelman S.J., Gulliver A.F., Kochukhov O.P., et al.} 2002, ApJ, 575, 449

\bibitem{auriere07}
{Auri{\`e}re M., Wade G.A., Silvester J., et al.} 2007, A\&A, 475, 1053

\bibitem{auriere09}
{{Auri{\`e}re} M., {Wade} G.A., {Konstantinova-Antova} R., et al.} 2009, A\&A, 504, 231


\bibitem{auriere10}
{{Auri{\`e}re} M., {Wade} G.A., {Ligni{\`e}res} F., et al.} 2010, A\&A, 523, A40 

\bibitem{babcock47}
{Babcock H.W.} 1947, ApJ, 105, 105


\bibitem{bagnulo02}
{{Bagnulo} S., {Szeifert} T., {Wade} G.A., et al.} 2002, A\&A, 389, 191

\bibitem{bagnulo09}
{{Bagnulo} S., {Landstreet} J.D., {Mason} E., et al.} 2009, A\&A, 450, 777

\bibitem{bagnulo12}
{{Bagnulo} S., {Landstreet} J.D., {Fossati} L., et al.} 2012, A\&A, 538, A129

\bibitem{bagnulo13}
{{Bagnulo} S., {Fossati} L., Kochukhov O., et al.} 2012, A\&A, in press (arXiv:1309.2158)


\bibitem{braithwaite06}
{Braithwaite J., Nordlund \AA.} 2006, A\&A, 450, 1077



\bibitem{donati97}
{{Donati} J.F., {Semel} M., {Carter} B.D., et al.} 1997, MNRAS, 291, 658

\bibitem{donati09}
{Donati J.F., Landstreet J.D.} 2009, ARA\&A, 47, 333

\bibitem{folsom10}
{{Folsom} C.P., {Kochukhov} O., {Wade} G.A., et al.} 2010, MNRAS, 407, 2383

\bibitem{hubrig98}
{Hubrig S.} 1998, CoSka, 27, 296

\bibitem{hubrig99}
{{Hubrig} S., {Castelli} F., {Wahlgren} G.M.} 1999, A\&A, 346, 139

\bibitem{hubrig01}
{{Hubrig} S., {Castelli} F.} 2001, A\&A, 375, 963

\bibitem{hubrig06a}
{{Hubrig} S., {Gonz\'alez} J.F., {Savanov} I., et al.} 2006, MNRAS, 371, 1953

\bibitem{hubrig06b}
{Hubrig S., North P., Sch\"oller M., Mathys G.} 2006, AN, 327, 289

\bibitem{hubrig09}
{{Hubrig} S., {Briquet} M., {De Cat} P., et al.} 2009, AN, 330, 317

\bibitem{hubrig09a}
{{Hubrig} S., {Sch\"oller} M., {Savanov} I., et al.} 2009, AN, 330, 708

\bibitem{hubrig11}
{{Hubrig} S., {Ilyin} I., {Sch\"oller} M., et al.} 2011, ApJ, 726, L5

\bibitem{hubrig12a}
{{Hubrig} S., {Sch\"oller} M., {Kholtygin} A.F., et al.} 2012, A\&A, 546, L6

\bibitem{hubrig12}
{{Hubrig} S., {Gonz\'alez} J.F., {Ilyin} I., et al.} 2012, A\&A, 547, A90

\bibitem{kochukhov05}
{{Kochukhov} O., {Piskunov} N., {Sachkov} M., et al.} 2005, A\&A, 439, 1093

\bibitem{kochukhov06}
{{Kochukhov} O., {Bagnulo} S.} 2006, A\&A, 450, 763

\bibitem{kochukhov07}
{{Kochukhov} O., {Adelman} S.J., {Gulliver} A.F., et al.} 2007, Nature Physics, 3, 526 

\bibitem{kochukhov10}
{{Kochukhov} O., {Makaganiuk} V., {Piskunov}, N.} 2010, A\&A, 524, A5

\bibitem{kochukhov11}
{{Kochukhov} O., {Makaganiuk} V., {Piskunov} N., et al.} 2011, A\&A, 534, L13

\bibitem{kochukhov13}
{{Kochukhov} O., {Makaganiuk} V., {Piskunov} N., et al.} 2013, A\&A, 554, A61

\bibitem{kochukhov13a}
{{Kochukhov} O., Sudnik N.} 2013, A\&A, 554, A93

\bibitem{kolenberg09}
{{Kolenberg} K., {Bagnulo} S.} 2009, A\&A, 498, 543

\bibitem{korhonen13}
{{Korhonen} H., {Gonz\'alez} J.F., {Briquet} M., et al.} 2013, A\&A, 553, A27


\bibitem{ligni09}
{{Ligni{\`e}res} F., {Petit} P., {B\"ohm} T., et al.} 2009, A\&A, 500, L41

\bibitem{maka11a}
{{Makaganiuk} V., {Kochukhov} O., {Piskunov} N., et al.} 2011, A\&A, 525, A97

\bibitem{maka11b}
{{Makaganiuk} V., {Kochukhov} O., {Piskunov} N., et al.} 2011, A\&A, 529, A160

\bibitem{maka12}
{{Makaganiuk} V., {Kochukhov} O., {Piskunov} N., et al.} 2012, A\&A, 539, A142

\bibitem{mathys94}
{Mathys G.} 1994, A\&AS, 108, 547

\bibitem{mathys95}
{{Mathys} G., {Hubrig} S.} 1995, A\&A, 293, 810

\bibitem{mathys06}
{{Mathys} G., {Hubrig} S.} 2006, A\&A, 453, 699

\bibitem{mikulasek08}
{{Mikul{\'a}{\v s}ek} Z., {Krti{\v c}ka} J., {Henry}, G.~W., et al.} 2008, A\&A, 485, 585

\bibitem{neiner12}
{{Neiner} C., {Alecian} E., {Briquet} M., et al.} 2012, A\&A, 537, A148

\bibitem{petit08}
{{Petit} P., {Dintrans} B., {Solanki} S.K., et al.} 2008, MNRAS, 388, 80

\bibitem{petit10}
{{Petit} P., {Ligni{\`e}res} F., Wade G.A., et al.} 2010, A\&A, 523, A41

\bibitem{petit11}
{{Petit} P., {Ligni{\`e}res} F., {Auri{\`e}re}  M., et al.} 2011, A\&A, 532, L13


\bibitem{sd12}
{{Sainz Dalda} A., {Mart{\'{\i}}nez-Sykora} J., {Bellot Rubio} L., et al.} 2012, ApJ, 748, 38

\bibitem{shorlin02}
{{Shorlin} S.L.S., {Wade} G.A., {Donati} J.F., et al.} 2002, A\&A, 392, 637

\bibitem{shultz12}
{{Shultz} M., {Wade} G.A., {Grunhut} J., et al.} 2012, ApJ, 750, 2

\bibitem{silvester09}
{{Silvester} J., {Neiner} C., {Henrichs} H.F., et al.} 2011, MNRAS, 398, 1505 

\bibitem{stibbs50}
{Stibbs D.W.N.} 1950, ApJ, 110, 395

\bibitem{wade06}
{{Wade} G.~A., {Auri{\`e}re} M., {Bagnulo} S., et al.} 2006, A\&A, 451, 293

\bibitem{wade12}
{{Wade} G.A., {Grunhut} J.H., {MiMeS Collaboration}} 2012, in ASP Conf. Ser. 464, 405

\end{thebibliography}
